\documentclass[runningheads]{llncs}

\usepackage[T1]{fontenc}
\usepackage{graphicx}
\usepackage{xspace}
\usepackage{hyperref}
\usepackage{tikz}
\usetikzlibrary{arrows}
\usepackage{tabularx}

\newcommand{\codeintext}[1]{\texttt{#1}}

\usepackage{listings, xcolor}

\definecolor{verylightgray}{rgb}{.97,.97,.97}
\definecolor{darkgreen}{rgb}{0,.5,0}

\lstdefinelanguage{Solidity}{
  basicstyle=\tiny\ttfamily, 
	keywords=[1]{anonymous, assembly, assert, balance, break, call, callcode, case, catch, class, constant, continue, contract, debugger, default, delegatecall, delete, do, else, emit, event, export, external, false, finally, for, function, gas, if, implements, import, in, indexed, instanceof, interface, internal, is, length, library, log0, log1, log2, log3, log4, memory, modifier, new, payable, pragma, private, protected, public, pure, push, require, return, returns, revert, selfdestruct, send, storage, struct, suicide, super, switch, then, this, throw, transfer, true, try, typeof, using, view, while, with, addmod, ecrecover, keccak256, mulmod, ripemd160, sha256, sha3}, 
	keywordstyle=[1]\color{blue}\bfseries,
	keywords=[2]{address, bool, byte, bytes, bytes1, bytes2, bytes3, bytes4, bytes5, bytes6, bytes7, bytes8, bytes9, bytes10, bytes11, bytes12, bytes13, bytes14, bytes15, bytes16, bytes17, bytes18, bytes19, bytes20, bytes21, bytes22, bytes23, bytes24, bytes25, bytes26, bytes27, bytes28, bytes29, bytes30, bytes31, bytes32, enum, int, int8, int16, int24, int32, int40, int48, int56, int64, int72, int80, int88, int96, int104, int112, int120, int128, int136, int144, int152, int160, int168, int176, int184, int192, int200, int208, int216, int224, int232, int240, int248, int256, mapping, string, uint, uint8, uint16, uint24, uint32, uint40, uint48, uint56, uint64, uint72, uint80, uint88, uint96, uint104, uint112, uint120, uint128, uint136, uint144, uint152, uint160, uint168, uint176, uint184, uint192, uint200, uint208, uint216, uint224, uint232, uint240, uint248, uint256, var, void, ether, finney, szabo, wei, days, hours, minutes, seconds, weeks, years},	
	keywordstyle=[2]\color{teal}\bfseries,
	keywords=[3]{block, blockhash, coinbase, difficulty, gaslimit, number, timestamp, msg, data, gas, sender, sig, value, now, tx, gasprice, origin},	
	keywordstyle=[3]\color{violet}\bfseries,
	identifierstyle=\color{black},
	sensitive=false,
	comment=[l]{//},
	morecomment=[s]{/*}{*/},
	commentstyle=\color{darkgreen}\ttfamily,
	stringstyle=\color{red}\ttfamily,
	morestring=[b]',
	morestring=[b]",
	numberstyle=\tiny,
	emphstyle=\underline
}

\lstdefinelanguage{Boogie}{
  basicstyle=\tiny\ttfamily, 
	keywords=[1]{axiom, break, call, const, else, exists, extends, forall, function, goto, if, implementation, modifies, old, procedure, returns, then, type, unique, var, while}, 
	keywordstyle=[1]\color{blue}\bfseries,
	keywords=[2]{address, bool, int, real}, 
	keywordstyle=[2]\color{teal}\bfseries,
	keywords=[3]{assert, assume, ensures, invariant, requires}, 
	keywordstyle=[3]\color{darkgreen}\bfseries,
	keywords=[4]{\_msg\_sender, \_msg\_value, \_balance, \_this}, 
	keywordstyle=[4]\color{violet}\bfseries,
	identifierstyle=\color{black},
	sensitive=false,
	comment=[l]{//},
	morecomment=[s]{/*}{*/},
	commentstyle=\color{gray}\ttfamily,
	stringstyle=\color{red}\ttfamily,
	morestring=[b]',
	morestring=[b]",
  numberstyle=\tiny
}

\newcommand{\addresstype}{address}
\newcommand{\boogiethis}{\_this}
\newcommand{\boogiemsgsender}{\_msg\_sender}
\newcommand{\boogiemsgvalue}{\_msg\_value}

\newcommand{\tool}{\textsc{solc-verify}\xspace}
\newcommand{\zzz}{\textsc{z3}\xspace}
\newcommand{\yices}{\textsc{yices2}\xspace}
\newcommand{\cvc}{\textsc{cvc4}\xspace}
\newcommand{\boogie}{\textsc{boogie}\xspace}

\newcommand{\mythril}{\textsc{mythril}\xspace}
\newcommand{\slither}{\textsc{slither}\xspace}

\newcommand\copyrighttext{%
	\footnotesize \textit{\textbf{Authors' manuscript.}}\quad Published in S. Chakraborty and J. A. Navas (Eds.): \textit{VSTTE 2019}, LNCS 12031, 2020. The final publication is available at Springer via {\url{https://doi.org/10.1007/978-3-030-41600-3_11}}. }
\newcommand\copyrightnotice{%
	\begin{tikzpicture}[remember picture,overlay]
	\node[anchor=south,yshift=2.5cm] at (current page.south) {\fbox{\parbox{\dimexpr\textwidth-\fboxsep-\fboxrule\relax}{\copyrighttext}}};
	\end{tikzpicture}%
}

\begin{document}
\title{\tool: A Modular Verifier for Solidity Smart Contracts}
%
%
\author{
\'{A}kos Hajdu\inst{1}%
\thanks{The author was also affiliated with SRI International as an intern during this project.}
\and
Dejan Jovanovi\'{c}\inst{2}
}
\authorrunning{\'{A}. Hajdu and D. Jovanovi\'{c}}
%
\institute{
Budapest University of Technology and Economics, Budapest, Hungary\\
\email{hajdua@mit.bme.hu}
\and
SRI International, New York City, USA\\
\email{dejan.jovanovic@sri.com}
}
\maketitle              
\begin{abstract}
We present \tool, a source-level verification tool for
Ethereum smart contracts. \tool takes smart contracts
written in Solidity and discharges verification conditions
using modular program analysis and SMT solvers. Built on top
of the Solidity compiler, \tool reasons at the level of the
contract source code, as opposed to the more common
approaches that operate at the level of Ethereum bytecode.
This enables \tool to effectively reason about
high-level contract properties while modeling low-level language
semantics precisely. The properties, such as contract
invariants, loop invariants, and function pre- and
post-conditions, can be provided as annotations in the code
by the developer. This enables automated, yet user-friendly
formal verification for smart contracts. We demonstrate
\tool by examining real-world examples where our tool can
effectively find bugs and prove correctness of non-trivial
properties with minimal user effort.%
\copyrightnotice%
\end{abstract}
%
%
%


\section{Introduction}

Distributed blockchain-based applications are gaining
traction as a secure and trustless alternative to more
centralized solutions that require trusted intermediaries
such as banks. The focus of early blockchain
implementations, such as Bitcoin \cite{nakamoto2008bitcoin},
was to provide the infrastructure for one particular
application: digital money (cryptocurrency). Public
blockchains allow arbitrary parties to transact with each
other in a secure and trustless manner, with no central
authority. In this setting a blockchain is a distributed
ledger of transactions, where nodes in a peer-to-peer
network are processing and validating transactions to
maintain integrity. The next step in the evolution of
blockchains was to extend the blockchain to a setting where
the digital money can also be programmable. This is achieved
by generalizing the ledger to allow deployment of programs
(termed smart contracts~\cite{szabo1994smart}) that operate
over ledger data. Blockchains with support for smart
contracts provide a general distributed computing platform
and allow a set of mutually distrusting parties to execute
and enforce their contractual terms (expressed as code)
automatically. At the moment, the most popular such platform
is the Ethereum
blockchain~\cite{wood2014ethereum,antonopoulus2018mastering}.

However, smart contracts are often prone to errors with
potentially devastating financial effects (see, e.g.,
\cite{atzei2017survey} for a survey). The infamous DAO
bug~\cite{dhillon2017dao} is an illustrative example of the
difficulties involved in deploying a smart contract. The DAO
was a relatively small contract (2KLOC of Solidity code)
that was heavily scrutinized by the wider Ethereum community
before deployment. Nevertheless, an attacker managed to
exploit a subtle reentrancy bug to steal \$60M worth of
cryptocurrency.
%
%
Examples such as the DAO highlight the mission-critical
nature of smart contracts. Although the code of the contract
is usually small by the standards of modern software, if the
contract attracts a large amount of investment, the code
carries a significant amount of value per line of code.
Moreover, since the contract code is stored on the
blockchain, once deployed, the code is immutable and making
upgrades or bug-fixes is impossible without complex
solutions that involve a central authority.
%
%
There has been a great interest in applying formal methods
to verify smart
contracts~\cite{atzei2017survey,harz2018safer,miller2018smart}.
While there are ongoing projects based on
identifying specific vulnerability
patterns~\cite{bhargavan2016formal,grishchenko2018semantic,tsankov2018securify,luu2016making,nikolic2018finding,mythril,slither},
theorem
provers~\cite{hildenbrandt2017kevm,hirai2017defining,sergey2018scilla},
finite
automata~\cite{mavridou2018fsolidm,abdellatif2018formal} or
SMT~\cite{leonardo2018smt,kalra2018zeus,lahiri2018formal}, they all have limitations in terms
of scalability, precision, expressiveness and ease of use.

In this paper we present \tool, a tool for formal
verification of Ethere\-um smart contracts that integrates
seamlessly with developer tools. \tool follows the modular
software verification approach (e.g., VCC
\cite{cohen2009vcc},
HA\-VOC~\cite{chatterjee2007reachability}, and
ESC/Java~\cite{flanagan2002extended}), in the context of
Solidity. Given a Solidity contract,
annotated with specifications, \tool translates the
contract into the Boogie intermediate verification
language~\cite{deline2005boogiepl,leino2008boogie}, and
discharges verification conditions by SMT
solvers~\cite{barrett2018satisfiability}.
Developers can define the expected behavior of their
contracts using annotations within the contract code,
including assertions, contract and loop invariants, and
function pre- and post-conditions. Verification of smart
contracts brings domain-specific challenges. To start
with, the semantics of Solidity include
Ethereum-specific constructs such as the blockchain state,
transactions, and data-types not common in general
programming languages. As an example, Ethereum smart
contracts generally operate on 256-bit integers, making
precise reasoning about low-level properties, such as the
absence of overflows, infeasible with standard SMT
techniques. Furthermore, some common high-level properties
of smart contracts, such as ``the sum of user balances is
always equal to the total supply'', cannot be expressed in
first-order logic or in Solidity, and therefore need
domain-specific treatment. \tool addresses these issues
through an SMT-friendly encoding of Solidity into Boogie
that is expressive enough to capture the properties of
interest, and takes advantage of recent advances in SMT
solving to enable effective reasoning. We describe \tool
through examples and
demonstrate how \tool can both find non-trivial bugs in
real-world examples and prove correctness after the bugs
have been fixed (e.g., the BEC token~\cite{bec2018cve} hack).
As far as we know, \tool is the first tool that allows specification
and modular verification of Solidity smart contracts that is
practical and automatic. \tool is implemented as an add-on
to the open-source Solidity compiler and is available on
GitHub.\footnote{\url{https://github.com/SRI-CSL/solidity}}


\section{Background}
\label{sec::background}


\paragraph{Ethereum.}

Ethereum~\cite{wood2014ethereum,antonopoulus2018mastering}
is a generic blockchain-based distributed computing
platform. The Ethereum ledger is a storage layer for a
database of \emph{accounts} and data associated with those
accounts, where each account is identified by its
\emph{address}. Ethereum contracts are usually written in a
high-level programming language, most notably
Solidity~\cite{soliditydoc}, and then compiled into the
bytecode of the Ethereum Virtual Machine (EVM).
A compiled contract is deployed to the blockchain
using a special transaction that carries the contract code
and sets up the initial state with the constructor.
At that point the deployed contract is
issued an address and stored on the ledger. From then on,
the contract is publicly accessible and its code cannot be
modified. A user (or another contract) can interact with a
contract through its public API by calling public
functions. This can be done by issuing a \emph{transaction} with the
contract's address as the recipient. The transaction
contains the function to be called along with the
arguments, and an execution fee called \emph{gas}.
Optionally, some value of Ether (the native currency of
Ethereum) can also be transferred with
transactions. The Ethereum network then executes the
transaction by running the contract code in the context of
the contract instance. During their execution, each
instruction costs some predefined amount of gas. If the
contract overspends its gas limit, or there is a runtime
error (e.g., an exception is thrown, or an assertion is
triggered), the entire transaction is aborted and has no
effect on the ledger (apart from charging the sender for the
used gas).

\paragraph{Solidity.}

Figure~\ref{fig:simplebank} shows a Solidity contract
\texttt{SimpleBank} that illustrates some of the common
features that Ethereum contracts use in practice.
A contract can have \emph{state variables}, which define the
persistent data that the contract will store on the ledger.
The state of \texttt{SimpleBank} consists of a
single variable \codeintext{balances}, which is a
\emph{mapping} from addresses to 256-bit integers. Further
Solidity types include \emph{value types}, such as Booleans,
signed and unsigned integers (of various bit-lengths),
addresses, fixed-size arrays, enums, and \emph{reference
types}, to be used with arbitrary-size arrays and
structures. Once deployed, an instance of
\texttt{SimpleBank} will be assigned its address and since
no constructor is provided, its data will be initialized to
default values (in this case an empty mapping).


\begin{figure}[htb]
\begin{minipage}[t]{.5\linewidth}
\begin{lstlisting}[language=Solidity,escapechar=|]
/** @notice invariant sum(balances) == this.balance */|\label{bank::invar}|
contract SimpleBank {
	mapping(address => uint256) balances;

	function deposit() payable public {
		balances[msg.sender] += msg.value;
	}

	function withdraw(uint256 amount) public {
		require(balances[msg.sender] > amount);
		if (!msg.sender.call.value(amount)("")) { |\label{bank::send}|
			revert();
		}
		balances[msg.sender] -= amount; |\label{bank::subtract}|
	}
}
\end{lstlisting}
\caption{
An example Solidity smart contract implementing a simple
bank. Users can deposit and withdraw Ether with the
corresponding functions, and the contract keeps track of
user balances. The top level annotation states that the
contract will ensure that the sum of individual balances is
equal to the total balance in the bank.
}
\label{fig:simplebank}
\end{minipage}
\hspace{0.05\linewidth}
\begin{minipage}[t]{.45\linewidth}
\begin{lstlisting}[language=Solidity,escapechar=|]
/** @notice invariant x == y */|\label{annot::invar}|
contract C {
	int x;
	int y;

	/** @notice precondition x == y|\label{annot::pre}|
	    @notice postcondition x == (y + n) */|\label{annot::post}|
	function add_to_x(int n) internal {
		x = x + n;|\label{annot::overflow}|
		require(x >= y); // Catch overflow|\label{annot::overflowcheck}|
	}

	function add(int n) public {
		require(n >= 0);
		add_to_x(n);
		/** @notice invariant y <= x */|\label{annot::loop}|
		while (y < x) {
			y = y + 1;
		}
	}
}
\end{lstlisting}
\caption{
An example Solidity smart contract illustrating the
annotation features of \tool, including contract-level
invariants, pre- and postconditions and loop invariants.
}
\label{fig:annotations}
\end{minipage}
\end{figure}

Contracts define \emph{functions} that can act on their state.
Functions can receive data as arguments,
perform computation, manipulate the state variables and
interact with other accounts. In addition to declared
parameters, functions also receive a \codeintext{msg}
structure that contains the details of the transaction. Our
example contract defines two public functions
\codeintext{deposit} and \codeintext{withdraw}. The
\codeintext{deposit} function is marked as
\codeintext{public} and \codeintext{pay\-ab\-le}, meaning
that it can be called by anyone and is allowed to receive
Ether as part of the call. This function reads the amount of
Ether received from \codeintext{msg.value} and adds it to
the balance of the caller, whose address is available in
\codeintext{msg.sender}. The \codeintext{withdraw} function
allows users to withdraw a part of their bank balance. The
function first checks that the sender's balance in the bank
is sufficient using a \codeintext{require} statement. If the
condition of \codeintext{require} fails, the transaction is
reverted with no effect. Otherwise the function sends the
required amount of Ether funds by using a \codeintext{call}
on the caller address with no arguments (denoted by the
empty string). The amount to be transferred is set with the
\codeintext{value} function. The recipient of the
\codeintext{call} can be another contract that can perform
arbitrary actions on its own (within the gas limits) and can
also fail (indicating it in the return value). If
\codeintext{call} fails, the whole transaction is reverted
with an explicit \codeintext{revert}, otherwise the balance
of the caller is deducted in the mapping as well.

\codeintext{SimpleBank} contains a classic
reentrancy vulnerability that can be exploited to steal
funds from the bank. As the control is transferred to the
caller in line~\ref{bank::send}, before their balance is
deducted in line~\ref{bank::subtract}, they are free to make
another call to \codeintext{withdraw} to perform a double
(or multiple) spend. Although this flaw seems basic, it
is the issue that lead to the loss of funds in the DAO hack
\cite{dhillon2017dao}.


\section{Overview and Features}


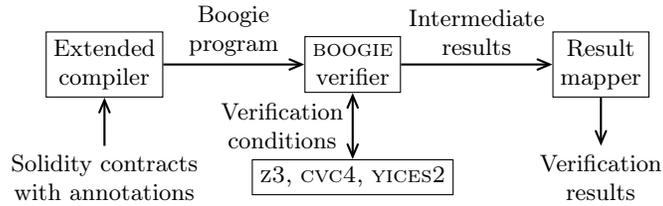
\begin{figure}[t]
	\centering
	\begin{tikzpicture}
	\tikzstyle{cnt}=[align=center]
	\tikzstyle{small}=[font=\small]	\tikzstyle{box}=[rectangle,align=center,draw,fill=white,inner sep=0.1cm]
	\tikzstyle{arr}=[-angle 45,thick]

	\node[box] (comp)  at (-3.3,  0  ) {Extended\\compiler};
	\node[box] (verif) at ( 0  ,  0  ) {\boogie\\verifier};
	\node[box] (map)   at ( 3.3,  0  ) {Result\\mapper};
	\node[box] (smt)   at ( 0  , -1.5) {\zzz, \cvc, \yices};
	\node[cnt] (inp)   at (-3.3, -1.5) {Solidity contracts\\with annotations};
	\node[cnt] (outp)  at ( 3.3, -1.5) {Verification\\results};

	\draw[arr] (inp)--(comp);
	\draw[arr] (comp)--(verif) node[midway,above,cnt,small]{Boogie\\program};
	\draw[arr] (verif)--(map) node[midway,above,cnt,small]{Intermediate\\results};
	\draw[arr] (map)--(outp);
	\draw[angle 45-angle 45,thick] (verif)--(smt) node[midway,left,cnt,small]{Verification\\conditions};
	\end{tikzpicture}
	\caption{Overview of the \tool modules. The extended compiler creates a Boogie program from the Solidity contract, which is checked by the \boogie verifier using SMT solvers. Finally, results are mapped back and presented at the Solidity code level.}
	\label{fig:imploverview}
\end{figure}

\tool is implemented as an extension to the Solidity compiler.
It takes
a set of Solidity contracts including specification annotations
and discharges verification conditions using the Boogie verifier
and SMT solvers. An overview of the architecture is
shown in Figure~\ref{fig:imploverview}.

\paragraph{Specification.}

Solidity provides only a few error handling constructs (e.g.,
\codeintext{assert}, \codeintext{require}) for the programmer to
specify expected behavior.
Therefore, \tool supports in-code \emph{annotations} to
specify contract properties, as illustrated in
Figure~\ref{fig:annotations}.
Annotations are side-effect free Solidity expressions, which
can reference any variable in the scope of the annotated element.
\emph{Contract-level invariants} (line~\ref{annot::invar})
must hold before and after the execution of every public
function and after the contract constructor.
Non-public functions are inlined to a depth of one by default, but
can also be specified with \emph{pre- and postconditions}
(lines~\ref{annot::pre}--\ref{annot::post}). Moreover,
\emph{loop invariants} (line~\ref{annot::loop}) can be
attached to loops. As an extension, we also provide a special \emph{sum}
function over collections (arrays and mappings) in
the specification language, as seen for example for
\codeintext{SimpleBank} in Figure~\ref{fig:simplebank}.
The sum function is modeled internally by associating a ghost variable
to the collection tracked by the sum: each collection
update also updates the ghost variable. This encoding is a
sufficient abstraction for our needs.

\paragraph{Correctness.}

\tool targets functional correctness of contracts with respect to completed\footnote{Due to the usage of gas, total and partial correctness are equivalent. Furthermore, currently we do not model gas: running out of gas does not affect correctness as the transaction is reverted. However, we might model it in the future in order to verify liveness properties or to be able to specify an upper bound.} transactions and different types of failures.
An \emph{expected failure} is a failure due to an exception deliberately thrown to guard from the user (e.g., \codeintext{require}, \codeintext{revert}).
An \emph{unexpected failure} is any other failure (e.g., \codeintext{assert}, overflow).
We say that a contract is \emph{correct} if all transactions
(public function calls) that do not fail due to an
expected failure also do not fail due to an unexpected failure and satisfy their specification.

\paragraph{Translation to Boogie.}

\tool relies on the Solidity compiler
that parses the contracts and builds an abstract syntax tree (AST)
where names, references and types are resolved. \tool then
traverses the internal AST and produces a
Boogie~\cite{deline2005boogiepl,leino2008boogie}
representation of the program. We discuss the details and
properties of the translation in more detail in
Section~\ref{sec:translation}.

\paragraph{Boogie and SMT.}

Boogie transforms the program into verification conditions (VCs)
and discharges them using SMT solvers. By default, Boogie
can use \zzz~\cite{demoura2008z3} and
\cvc~\cite{barrett2011cvc4} but we also extended it to
support \yices~\cite{dutertre2014yices}. A notable
feature of our encoding is that it allows quantifier-free VC
generation, permitting to use SMT solvers that do not
support quantifiers (e.g., \yices). Boogie reports violated
annotations and failing assertions in the Boogie program
and \tool maps these errors back to the
Solidity code using traceability information. The final
output of \tool is a list of errors corresponding to
the original contracts (e.g., line numbers, function names).

\section{Translation Details and Properties}
\label{sec:translation}
The core of \tool is a translation from Solidity
contracts to the Boogie IVL, supporting a majority of
the Solidity language.%
\footnote{The paper and the experiments are based on compiler
	version v0.4.25, but we keep \tool up to date with the
	latest development branch.}

\paragraph{Contracts.}
The input of the translation is a collection of contracts
to be verified and the output is a single Boogie program with
all contracts.
\tool can reason about single and multiple contracts as well.
If the code of all contracts is available, \tool can take all
available annotations into account when reasoning. However, this
can be unsafe as EVM addresses are not typed (any address can be
cast to a contract type) and is to be used with care.
\tool also supports inheritance by relying on the compiler to
perform flattening and virtual-call disambiguation.

\paragraph{Types.}

\tool supports basic Solidity types such as Booleans, integers
and addresses.
Several modes are provided for modeling arithmetic
operations that can be selected by the user. In the simplest
mode, integers are unbounded \emph{mathematical integers}.
This mode does not capture the exact semantics of the
operations (e.g., overflows) but is scalable and well
supported by SMT solvers. Precise arithmetic can be provided
by relying on the SMT theory of \emph{bitvectors}. \tool
supports this mode but can suffer from scalability issues
due to the 256-bit default integer size of Solidity. In
order to provide both precision and scalability, \tool
provides a \emph{modular arithmetic} mode that encodes
arithmetic operations using mathematical integers with range
assertions and precise wraparound semantics of all
operations.
Addresses are modeled with uninterpreted
symbols as they can only be queried for equivalence.
\tool also supports mappings and arrays using SMT
arrays~\cite{mccarty1962arrays,de2009generalized}.
Structures, enumerations and tuples are currently not
supported but there are no technical difficulties in supporting
them and they are planned in the future.
Events (a logging mechanism)
are
ignored as they are not relevant for functional correctness.\footnote{We might model events in the future to be able to specify that an event is expected to be triggered.}

\paragraph{State Variables.}
State variables are mapped to global variables in Boogie.
However, multiple instances of a contract can be deployed
to the blockchain at different addresses. Since aliasing of contract storage is
not possible, \tool models each state variable as a
one-dimensional global mapping from contract addresses
to their respective type (in essence treating the blockchain
as a heap in a Burstall-Bornat model~\cite{bornat2000proving}).
For example, the state variable \codeintext{x}
with type \codeintext{int} at line~\ref{line:statevar} of
Figure~\ref{fig:basics} (left) is transformed to the global
variable \codeintext{x} with mapping type \codeintext{[address]int}
at line~\ref{line:globalvar} of
Figure~\ref{fig:basics} (right).

\begin{figure*}[htb]

\begin{minipage}{.45\linewidth}%
\begin{lstlisting}[language=Solidity,escapechar=|]
contract A {
	int public x; |\label{line:statevar}|
	function set(int _x) public { x = _x; }
}
contract B {
	A a;
	function setXofA(uint x) public { a.set(x); }
	function getXofA() public returns (uint) {
		return a.x();
	}
}
\end{lstlisting}
\end{minipage}
\hfill
\begin{minipage}{.49\linewidth}%
\begin{lstlisting}[language=Boogie,escapechar=|]
var x: [address]int; |\label{line:globalvar}|
procedure set(_this: address, _x: int) {
	x := x[_this :=_x];
}
var a: [address]address;
procedure setXofA(_this: address, x: int) {
	call set(a[_this], x);
}
procedure getXofA(_this: address) returns (r: int) {
	r := x[a[_this]];
}
\end{lstlisting}
\end{minipage}
\caption{Solidity contract (left) and its Boogie translation (right), illustrating the representation of the blockchain data as a heap and the receiver parameter of functions.}
\label{fig:basics}
\end{figure*}

\paragraph{Functions.}
Each function in Solidity is translated to a procedure in
Boogie with an
additional implicit receiver
parameter~\cite{barnett2004verification} called
\codeintext{\boogiethis}, which identifies
the address of the contract instance. As an example,
consider the \codeintext{set} function of the Solidity
contract \codeintext{A} in Figure~\ref{fig:basics}.
Updating \codeintext{x} in the Boogie program
becomes an update of the map \codeintext{x} using the
receiver parameter \codeintext{\boogiethis}. Consider also
the call \codeintext{a.set(x)} in the Solidity function
\codeintext{setXofA}. The Boogie program first gets the
address of the \codeintext{A} instance corresponding to the
current \codeintext{B} instance using
\codeintext{a[\boogiethis]}. Then it passes this address to
the receiver parameter of the function
\codeintext{set}.

Functions can be declared view (cannot write state)
or pure (cannot read or write state), but these restrictions are
checked by the compiler.
Additional user-defined function modifiers
are a language feature of Solidity to alter or extend the
behavior of functions. In practice, modifiers are commonly
used to weave in extra checks and instructions to
functions.
For example, the \codeintext{pay} function in
Figure~\ref{fig:globalstuff} (left) includes the modifier
\codeintext{onlyOwner} (defined in
line~\ref{wallet::solidity_modifier}), which performs an extra
check before calling the actual function (denoted by the
placeholder \codeintext{\_}).
\tool simply inlines statements of all modifiers
of a function to obtain a single Boogie procedure (e.g.,
\codeintext{pay} procedure in Figure~\ref{fig:globalstuff} right).

\begin{figure*}
\begin{minipage}{.37\linewidth}%
\begin{lstlisting}[language=Solidity,escapechar=|]
contract Wallet {
	address owner;
	
	modifier onlyOwner() { |\label{wallet::solidity_modifier}|
		require(msg.sender == owner); |\label{wallet::solidity_sender}|
		_;
	}
	function receive() payable public { |\label{wallet::solidity_payable}|
		// Actions could be performed here
	}
	function pay(address to, uint amount) public onlyOwner {
		to.transfer(amount); |\label{wallet::solidity_transfer}|
	}
}
\end{lstlisting}
\end{minipage}
\hfill
\begin{minipage}{.58\linewidth}%
\begin{lstlisting}[language=Boogie,escapechar=|]
var _balance: [address]int; |\label{wallet::boogie_balance}|

var owner: [address]address;

procedure receive(_this: address, _msg_sender: address, _msg_value: int) {
	_balance := _balance[_this := _balance[_this] + _msg_value]; |\label{wallet::boogie_payable}|
	// Actions could be performed here
}
procedure pay(_this: address, _msg_sender: address, _msg_value: int, to: address, amount: int) {
	assume(_msg_sender == owner[_this]); |\label{wallet::boogie_modifier}|
	assume( _balance[_this] >= amount); |\label{wallet::boogie_transfer1}|
	_balance := _balance[_this := _balance[_this] - amount];
	_balance := _balance[to := _balance[to] + amount]; |\label{wallet::boogie_transfer2}|
}
\end{lstlisting}
\end{minipage}
\caption{A simple wallet, which can receive Ether from anyone but only the owner can make transfers. This example illustrates various Ethereum and blockchain features in Solidity (left) along with their representation in Boogie (right).}
\label{fig:globalstuff}
\end{figure*}

\paragraph{Statements and expressions.}
Most of the Solidity statements and expressions can be
directly mapped to a corresponding statement or expression
in Boogie with the same semantics, including variable
declarations, conditionals, \codeintext{while} loops, calls,
returns, indexing, unary/binary operations and literals.
There are also some statements and expressions that require
a simple transformation, such as mapping \codeintext{for}
loops to \codeintext{while} loops or extracting nested calls
and assignments within expressions to separate statements
using fresh temporary variables.
\tool currently does not support inline assembly and creating
new contracts from within another contract
(\codeintext{new} expressions).
Furthermore, the availability of some arithmetic operations
depends on the expressiveness of the underlying domain
(e.g., bitwise operations).

\paragraph{Transactions.}
Solidity includes Ethereum-specific functions and variables
to query and manipulate balances and transactions.
Some examples can be seen in Figure~\ref{fig:globalstuff}
(left) with the corresponding translation in
Figure~\ref{fig:globalstuff} (right).
Each address is associated with its balance, which can be queried using the \codeintext{balance}
member of the address. Correspondingly, \tool
keeps track of the balances in a global mapping from
addresses to integers (line~\ref{wallet::boogie_balance}
of Figure~\ref{fig:globalstuff} right).

Solidity offers the \codeintext{msg.sender} field within
functions (line~\ref{wallet::solidity_sender} of
Figure~\ref{fig:globalstuff} left) to
access the caller address. \tool maps this to Boogie by adding an
extra parameter \codeintext{\boogiemsgsender} of type
\codeintext{\addresstype} to each procedure. When a
procedure calls another, the current receiver address
(\codeintext{\boogiethis}) is passed in as the sender.

Solidity functions marked with the \codeintext{payable}
keyword (line~\ref{wallet::solidity_payable} of
Figure~\ref{fig:globalstuff} left) are capable of receiving
Ether when called. The amount of Ether received can be
queried from the
\codeintext{msg.value} field. \tool models this in Boogie by
including an extra parameter \codeintext{\boogiemsgvalue}
and updating the global balances
map at the beginning of the corresponding Boogie procedure
(line~\ref{wallet::boogie_payable} of
Figure~\ref{fig:globalstuff} right).
When calling a payable function in Solidity,
the amount of
Ether to be transferred can be set with the special
\codeintext{value} function (e.g., line~\ref{bank::send} of
Figure~\ref{fig:simplebank}). \tool translates this to Boogie by
reducing the balance of the caller before making the call
and passing the value as the \codeintext{\boogiemsgvalue}
argument.

The functions \codeintext{send} and \codeintext{transfer} are
dedicated functions to transfer Ether between addresses. \tool
inlines these functions by manipulating the global mapping of balances
directly. If the recipient is a contract, a special fallback
function is executed, but the gas passed is limited to raising events,
which is irrelevant for functional correctness.\footnote{Gas costs
of certain write operations were about to change with Constantinople,
allowing a reentrancy attack, but it was reverted with the St.~Petersburg
upgrade~\cite{ethereumblog2019constantinople}.}
For example, the transfer in
line~\ref{wallet::solidity_transfer} of Figure~\ref{fig:globalstuff}
(left) is mapped to lines~\ref{wallet::boogie_transfer1}--\ref{wallet::boogie_transfer2}
on the right. The sender not having enough funds is an expected
transaction failure, which is modeled with an assumption.

The function \codeintext{call} can call a function by its name on
any address and can also pass arbitrary data.
Since there can be an unknown code behind
the called address, \tool treats such cases as an external call
that can perform arbitrary computation.\footnote{Contract invariants
are also checked before external calls as they can perform
a callback to the contract.}
\tool does not support low-level function calls such as
\codeintext{callcode} and \codeintext{delegatecall}
as it is considered dangerous and would require encoding
of the EVM details (contract layout, EVM semantics).

\paragraph{Error handling.}
Solidity exceptions will undo all changes made to the global
state by the current call.
Deliberately thrown exceptions (\codeintext{require},
\codeintext{revert}, \codeintext{throw}) are therefore mapped
to assumptions in Boogie, which stop the verifier without
reporting an error.
Assertions are mapped to Boogie assertions, causing a reported
error when their condition evaluates to false.

\paragraph{Detection of overflows.}
Neither the EVM nor Solidity performs any checking
of the results of arithmetic operations by default. Due to
the wraparound semantics of integers, this allows unexpected
overflows and underflows to occur undetected (e.g., the
infamous BEC token \cite{bec2018cve}).

In general, overflows can be detected by checking the
result of every operation after it has been computed.
However, reporting every such overflow would
result in an overwhelming number of false alarms. For
example, it is common practice for Solidity developers to
perform arithmetic operations first, and then check for
overflows manually after the fact (see, e.g.,
line~\ref{annot::overflowcheck} of
Figure~\ref{fig:annotations}). This practice of overflow
detection is an integral part of the SafeMath
library~\cite{safemath}
that is used in almost all deployed contracts on the
Ethereum blockchain and is part of Solidity best practices
\cite{consensys2018bestpractices}.

To reduce the number of false overflow reports, \tool uses the
following approach. Whenever an arithmetic computation is
performed, it computes the \emph{overflow condition} that
captures whether the overflow has occurred (i.e., if the
result of the computation in modular arithmetic is different
from the result over
unbounded integers). However, instead of immediately
checking this condition, it is accumulated in a dedicated
Boolean overflow-detection variable. \tool then checks for
overflow at the end of every basic block with an assertion.
This ``delayed checking'' gives space to developer to
perform manual checking for the overflow (in which case the
assertion will not trigger) and will avoid the false alarms.
For example, the potential overflow
in line~\ref{annot::overflow} of
Figure~\ref{fig:annotations} is not reported because in the
very next line the programmer guards the overflow and
reverts the transaction.


\section{Examples and Experiments on Real World Contracts}
\label{sec:results}

To demonstrate \tool we first discuss the coverage of
currently-supported language features and scalability
by examining the (unannotated) contracts currently
deployed on the Ethereum blockchain. We also pick a subset
of the unannotated contracts and manually check what \tool
can report on them. Finally, we examine two contracts that
had been exploited in the past, and show how \tool could
have found the issues, with minimal annotation burden, and
prove that the fixed versions of the contracts are correct.

\subsection{Language Coverage}

To analyze the coverage of currently-supported language
features and the scalability of \tool, we collected 37531
contracts available on Etherscan.\footnote{
\url{http://csl.sri.com/users/dejan/contracts.tar.gz}}
These contracts were
compiled with various versions of the Solidity compiler and
not all of them are supported by version 0.4.25 that
\tool used at the time of writing the paper. We therefore
selected 7836 contracts that do compile. The results of
running \tool on the selected contracts is shown in
Table~\ref{tbl:etherscan}. Columns correspond to different
arithmetic modes, with the last column representing modular
arithmetic with overflow checking enabled. The first row
shows that roughly 50\% of the contracts can be
successfully translated to Boogie in each mode. Contracts
that cannot be translated contain constructs not yet handled
by \tool, such as structures, enumerations or special
transaction and blockchain members. Some features (e.g.,
exponentiation) also depend on the arithmetic mode,
resulting in slight differences in feature coverage. The
remaining three rows show the number of contracts for which
\tool terminates within 10s with a given SMT solver as a
backend.
Note, that the effectiveness of the different SMT solvers
on this set
of contracts should be taken with a grain of salt. For
example, the bitvector encoding seems to be nearly as
efficient as modular arithmetic. However, this is because the
assertions in these contracts do not depend on
complex (e.g., nonlinear) arithmetic. With more complex
invariants, the bitvector encoding becomes infeasible for
reasoning, as we demonstrate it with the BEC token example
later in this section.
The takeaway of these results is that the average
execution time per contract is around a second, meaning that
\tool is applicable and effective for a significant amount
of real-world contracts, but scalability might depend on the
complexity of the properties.


\newcolumntype{b}{>{\hsize=.28\hsize}X}
\newcolumntype{s}{>{\centering\arraybackslash\hsize=.18\hsize}X}
\renewcommand{\arraystretch}{1.05}
\begin{table}[t]
	\caption{Etherscan results with different solvers and
	arithmetic encodings. Each cell represents the number of
	successfully processed contracts (of 7836 total) and the
	average execution time per contract.}
	\label{tbl:etherscan}
	\begin{tabularx}{\columnwidth}{|b||s|s|s|s|}
		\hline
		Encoding & \texttt{int} & \texttt{bv}    & \texttt{mod}   & \texttt{mod-overflow} \\
		\hline
		Translated & 4096  & 3919  & 3926  & 3926 \\
		\hline
		\cvc & 4090 (0.71s)  & 3837 (0.99s)  & 3921 (0.72s) & 3911 (0.79s) \\
		\yices & 3892 (1.15s) & 3854 (0.86s) & 3903 (0.75s) & 3859 (0.87s) \\
		\zzz & 3897 (1.24s) & 3831 (1.10s) & 3892 (0.87s) & 3894 (0.88s) \\
		\hline
	\end{tabularx}
\end{table}

\subsection{Unannotated Contracts}

The contracts available at Etherscan are \emph{not
annotated} and \tool can only consider
\codeintext{assert} and \codeintext{require} statements, and
overflows as implicit specification. Furthermore, the ground
truth about the contracts (whether they are
correct or not) is unknown. Nevertheless, we systematically
selected a subset of the contracts and manually checked the results
given by \tool.

We took all 3897 contracts that \tool could translate and process
with \zzz in integer mode. At the first glance we discovered
that a majority of the contracts (2754) use the popular
SafeMath library~\cite{safemath}, which has just recently
adopted the proper usage of \codeintext{assert} and
\codeintext{require}.\footnote{For discussion, see
\url{https://github.com/OpenZeppelin/openzeppelin-solidity/issues/1120}.}
We updated these contracts to properly guard against user input
with \codeintext{require} (instead of \codeintext{assert}).
Afterwards, we checked for \emph{assertion failures} and
\emph{overflows} using \tool.

\paragraph{Assertion checking.}

Surprisingly, only 88 contracts (out of the 3897) contain
assertions. \tool reported an error for 80 contracts, which
we all checked manually. Out of those errors, 78 are clearly
false alarms caused by a bad specification -- the developer wrote \codeintext{assert} where \codeintext{require} should have been used --
and fit into one of the following categories:
\begin{itemize}
	\item Enforcing input validity with assertion (e.g., input
	arrays are of equal size).
	\item Enforcing time locks with an assertion (e.g., \codeintext{now > 100}).
	\item Enforcing success of functions calls with an assertion (e.g. \codeintext{addr.call()}).
	\item Enforcing permissions with an assertion
	(e.g., checking \codeintext{msg.sender}).
	\item Enforcing correct result of arithmetic operations with an assertion.
\end{itemize}
As described in the Solidity documentation~\cite{soliditydoc}
\codeintext{assert} should only be used to check for internal
errors and invariants, and all cases highlighted above should use \codeintext{require} instead. After replacing the spurious assertions
with \codeintext{require}, \tool reports no false alarms.

The 2 reported errors worth discussing in more detail are illustrated in
Figure~\ref{fig:unannotated_assert}. The example on the left is a pre-sale
contract that accepts Ether until a sale cap is reached. The invariant of the
contract, i.e. that \codeintext{(raised <= max)} is enforced with a (stronger)
assertion at the beginning of function. It could be argued that this fits within
the mentioned prescribed usage for the \codeintext{assert} construct. However,
as \tool performs modular analysis, and nothing is assumed about the state
before a function call, it will report such an assertion as a potential
error. To fix this, the invariant \codeintext{(raised <= max)} should be
specified as a contract invariant, and \codeintext{require} should be used
to check the stronger precondition at function entry (followed by an \codeintext{assert} at the end of the function).

The example on the right is a token transfer function. The function checks
whether the sender has enough balance, and then it transfers the tokens to the
recipient. Finally, the assertion checks that no overflow has occurred using an
\codeintext{assert} statement on the result of the addition. As is, \tool
reports an error because increasing the balance of the recipient might overflow.
As argued above, if the purpose of the assertion is to guard against overflows
\codeintext{require} should be used instead. On the other hand, one could argue
that for fixed-cap tokens such an overflow should never occur since no address
can hold enough tokens to trigger the overflow. This assumption can be
explicitly specified, i.e., by stating a contract invariant
\codeintext{sum(balances) <= cap}. With this invariant,
\tool avoids the false alarm by inferring that overflow is no longer
possible.


\begin{figure*}[t]
\begin{minipage}{.48\linewidth}%
\begin{lstlisting}[language=Solidity,escapechar=|]
uint max    = 1000 ether;
uint raised = 0;

function() payable {
	assert(raised < max);
	require(msg.value != 0);
	require(raised + msg.value <= max);
	raised += msg.value;
}
\end{lstlisting}
\end{minipage}
\hfill
\begin{minipage}{.48\linewidth}%
\begin{lstlisting}[language=Solidity,escapechar=|]
mapping (address => uint) balances;

function transfer(address to, uint val) {
	require(balances[msg.sender] >= val);
	require(msg.sender != to);
	balances[msg.sender] = balances[msg.sender] - val;
	balances[to] = balances[to] + val;
	assert(balances[to] >= val);
}
\end{lstlisting}
\end{minipage}
\caption{Examples of failing assertions reported by \tool.}
\label{fig:unannotated_assert}
\end{figure*}

\paragraph{Overflow checking.}

We also checked for overflows and manually verified
the results for the 68 contracts (out of 3897) that have at least 100
transactions. \tool reports 33 alarms of which 29 are false and 4 can be
considered as real. All false alarms are due to implicit assumptions on the
magnitude of used numbers. There are 20 false alarms due to missing range
assumptions for array lengths causing false overflow alarms for loop counters.
For example, in a loop \codeintext{for (uint i = 0; i < array.length; i++) \{\}}
\tool reports that \codeintext{i++} might overflow. It is reasonable to assume
that array lengths remain small due to the gas costs associated with growing an
array. Other false alarms are caused by implicit assumptions on Ether balances
or time. For example, it is assumed that a counter for the total amount of Ether
received by a contract, or multiplying \codeintext{msg.value} by 20000 cannot
overflow because the amount of Ether is limited. Similarly, adding days or even
weeks to the current timestamp will not overflow any time soon. We plan to
include such implicit assumptions to a limited extent but, in general, it is
best if the developer explicitly specifies them. The four issues
found that could be considered real are the following:
\begin{itemize}
	\item A pre-sale contract sets the \codeintext{hardCap} in its constructor
	based on a \codeintext{cap} provided as argument with \codeintext{hardCap =
	cap*(10**18)}. Although the constructor is only called once by the deployer,
	providing a large cap can result in an unintentional overflow.
	\item A crowd-sale contract sets the unit cost based on the argument
	\codeintext{perEther} by calculating \codeintext{unitCost = 1 ether /
	(perEther*10**8)}. The problematic function is guarded so that it can only be called by the contract owner. Nevertheless, overflow can happen and
	can lead to an inconsistent unit price.
	\item A utility contract for mass distribution of tokens has a function to
	transfer an array of values to an array of recipients as a batch. The total
	amount transferred is kept accumulated in a contract counter and can overflow.
	However, as the counter is not used otherwise, the overflow might be benign.
	\item A food store contract first calculates the cost based on the bundles
	ordered, by computing \codeintext{cost = bundles * price}, where
	\codeintext{bundles} is provided by the caller. The function then checks
	if \codeintext{msg.value >= cost} holds, but this check can be bypassed
	with the overflow, opening the door for a potential exploit.
\end{itemize}

\subsection{Annotated Contracts}
\label{sec:results:annotated}

While \tool can find violations to implicit specifications
in unannotated contracts, its main target is to allow developers
to check custom, high-level properties by the means of annotations.
We demonstrate this by annotating two
contracts, finding bugs and proving the correctness of the
fixed versions.

\paragraph{Reentrancy detection (DAO).}

Reentrancy is a common source of vulnerabilities and the
cause of the infamous DAO bug~\cite{dhillon2017dao}. As
explained in Section~\ref{sec::background}, the \codeintext{SimpleBank}
contract presented in Figure~\ref{fig:simplebank} suffers
from the same reentrancy bug. Using \tool, the developer can
specify the consistency of the bank contract state through a
contract-level invariant, and \tool can detect the bug. For
example, we can annotate the contract with a property
\codeintext{sum(balances) == this.balance}. As the balance
of the contract is deducted before the external call, the
contract invariant is violated and \tool reports a (real)
error. However, if the user fixes the issue by first
reducing the balance of the recipient in the mapping and
then transferring the amount, the invariant will hold before
making the external call and \tool proves the specification
successfully. For both the buggy and correct versions of the
contract, the verification with \tool is instant.

\paragraph{Overflow detection (BEC token).}


\begin{figure*}[t]
\begin{lstlisting}[language=Solidity,escapechar=§]
library SafeMath {
  function mul(uint256 a, uint256 b) internal pure returns (uint256) { §\label{bec::safemath_mul}§
    uint256 c = a * b;
    require(a == 0 || c / a == b);
    return c;
  }
  // Similar for add, sub, div
}

/** @notice invariant totalSupply == sum(balances) */ §\label{bec::invar}§
contract BecToken {
  using SafeMath for uint256;

	uint256 public totalSupply;§\label{bec::totalsupply}§
  mapping(address => uint256) balances;

  function batchTransfer(address[] _receivers, uint256 _value) public returns (bool) {
    uint cnt = _receivers.length;
    uint256 amount = uint256(cnt) * _value; // Overflow §\label{bec::overflow}§
    // uint256 amount = uint256(cnt).mul(_value); // Correct version §\label{bec::correction}§
    require(cnt > 0 && cnt <= 20);
    require(_value > 0 && balances[msg.sender] >= amount);
    balances[msg.sender] = balances[msg.sender].sub(amount);
    /** @notice invariant totalSupply == sum(balances) + (cnt - i) * _value §\label{bec::loopinvar1}§
        @notice invariant (i <= cnt) */
    for (uint i = 0; i < cnt; i++) {
      balances[_receivers[i]] = balances[_receivers[i]].add(_value);
    }
    return true;
  }
}\end{lstlisting}
\caption{
Annotated part of the BECToken contract relevant for
the ``batchOverflow'' bug~\cite{bec2018cve}. While the contract uses the
\codeintext{SafeMath} library for most of its operations,
there is a multiplication in line~\ref{bec::overflow} that
can overflow.
}
\label{fig:bectoken}
\end{figure*}

We now consider the BEC token vulnerability~\cite{bec2018cve}
that has been exploited and resulted in significant
financial losses. The relevant part of the contract is shown
in Figure~\ref{fig:bectoken}. The contract is a typical
token contract, tracking balances of users in terms of their
BEC tokens and allowing transfers of tokens between users.
The function \codeintext{batchTransfer} shown in the figure
is intended to be used for transferring some value of BEC
tokens to a group of recipients in a batch. To do so, the
contract multiplies the requested value with the number
of recipients. Unfortunately, this multiplication can result
in an overflow (line~\ref{bec::overflow}), causing the total
transfer amount to be invalid (e.g.,~0). This allows
attackers to ``print'' large amounts of tokens and send them
to other users, while keeping their own balance constant.
Running \tool with the modular encoding of arithmetic
successfully detects the overflow issue of BEC token and does
not report any other potential overflows. After fixing the
contract (line~\ref{bec::correction}), \tool shows that no overflows are possible.
We also annotated the BEC contract with a specification that
the contract maintains the correct token balances throughout
the operation. As before, we add the invariant
\codeintext{totalSupply == sum(balances)} to the contract,
and adapt it to the loop invariant. The loop invariant
introduces extra complexity as it involves nonlinear
arithmetic and illustrates the need for precise reasoning at
large bit-sizes. Running \tool on the annotated contract in
the bitvector mode does not terminate regardless of
the SMT solver used.\footnote{With bit-size of 16 bits, \zzz
can discharge the VCs in 2295s while other solvers do not
terminate.} On the other hand, using modular arithmetic with
overflow detection \tool discharges all VCs
(with 256-bit integers) in seconds for both the buggy and correct
version of the contract (with \cvc).

\paragraph{Other tools.}

As far as we know, \tool is the only available tool that can
reason effectively and precisely about Solidity code with
specifications. The Solidity compiler includes an experimental
SMT checker~\cite{leonardo2018smt}, which is currently limited
to basic require/assert and overflow checking.
For the BEC token the latest version (v0.5.10) reports every arithmetic operation as a
potential overflow, including all false alarms in the
\codeintext{SafeMath} library.
It cannot detect the reentrancy issue in the SimpleBank example
because external calls and the \codeintext{revert} function is
not supported.
Furthermore, it incorrectly reports that the condition for
revert is always true (possibly because \codeintext{call} is
skipped and the default return value is false).
\textsc{Zeus}~\cite{kalra2018zeus} is not available publicly
for comparison.\footnote{We could only obtain a spreadsheet of results from the authors.}
\textsc{VeriSol}~\cite{lahiri2018formal} does not support
libraries (like \codeintext{SafeMath}) or the \codeintext{call}
function, which can cause reentrancy so we could not apply
it to our examples.

Two notable static analysis tools are \mythril~\cite{mythril} and
\slither~\cite{slither}. \mythril (v0.20.0) correctly reports the
overflow issue with
the BEC token in 200s, but it also reports all spurious
overflows. \mythril detects the reentrancy issue with the
bank contract, but it also reports the same issue with the
corrected version of the contract. \slither (v0.5.2), on the other
hand, has a dedicated DAO-like reentrancy issue check and
correctly handles both the buggy and correct version of the
bank contract. However, \slither doesn't support overflow
checking and therefore doesn't detect the BEC token issue.

Our goal, as demonstrated by the annotated examples, is to
provide a tool that allows developers to check their own
high-level annotations and business logic properties.
This makes \tool a good complementary to other
automated verification tools
that mainly target well known vulnerability patterns.


\section{Related Work}

%

The popularity of blockchain technology and many high-profile attacks and
vulnerabilities have put focus on the need for formal verification for smart
contracts~\cite{atzei2017survey,harz2018safer,miller2018smart}. We mention
prominent advances relying on vulnerability patterns, theorem provers, finite
automata and SMT, and relate them to our work.

\paragraph{Vulnerability pattern-based approaches.}

Bhargavan et~al.~\cite{bhargavan2016formal} translate a fragment of Solidity and
EVM to F$^*$ and use its type and effect system to check for vulnerable patterns
and gas boundedness. Grishchenko et~al.~\cite{grishchenko2018semantic} extend
this work on EVM by
checking security properties such as call integrity, atomicity, and independence
from miner controlled parameters. \textsc{Securify}~\cite{tsankov2018securify}
decompiles EVM and infers data- and control-flow dependencies in Datalog to
check for compliance and violation patterns.
\textsc{Oyente}~\cite{luu2016making} is a symbolic execution tool that can check
various patterns, including transaction ordering dependency, timestamp
dependency, mishandled exceptions and reentrancy.
\textsc{Maian}~\cite{nikolic2018finding} uses symbolic analysis with concrete
validation over a sequence of invocations to detect fund locking, fund leaking
and contracts that can be killed. \mythril~\cite{mythril} uses symbolic analysis
to detect a variety of security vulnerabilities. \slither~\cite{slither} is a
static analysis framework with dedicated vulnerability checkers. Approaches
based on vulnerability patterns, as the ones mentioned above, can be effective
at discharging specific properties, but are
limited to built-in patterns (or a domain specific
language~\cite{tsankov2018securify}). Furthermore, as they are mainly
EVM-based it makes reasoning about more general properties difficult. Our approach
focuses on Solidity and allows high-level, user-defined properties to
be checked effectively.

\paragraph{Theorem prover-based approaches.}

\textsc{Kevm}~\cite{hildenbrandt2017kevm} is an executable formal semantics of
EVM based on the K~framework including a deductive
program verifier to check contracts against given specifications. Hirai~\cite{hirai2017defining}
formalizes EVM in Lem, a language used by various
theorem provers and proves properties using interactive theorem proving.
Scilla~\cite{sergey2018scilla} is an intermediate language between smart
contracts and bytecode, using the Coq proof assistant for reasoning. Theorem
prover-based approaches offer the ability to capture precise, formal semantics
of the contracts but can be cumbersome as properties also need to be
formalized in the language of the theorem prover. Moreover, user interaction and
assistance is usually required impeding usability for contract
developers.\footnote{For an example of the difficulties in manually analyzing
even trivial issues, see \url{https://runtimeverification.com/blog/erc-20-verification/}.} In
our approach the developer can specify the properties directly within the
contract, as Solidity annotations and modular verification is fully automated.
Although loop invariants might be required, complex loops are rare in
contracts.


\paragraph{Automata-based approaches.}

\textsc{FSolidM}~\cite{mavridou2018fsolidm} is a finite state machine-based
designer for smart contracts that can generate Solidity code. Security features
and design patterns (e.g., locking, access control) can be included in the state
machine. Abdellatif and Brousmiche~\cite{abdellatif2018formal} model contracts
and the blockchain manually in BIP and use statistical model checking to
simulate uncertainties in the environment. Such model-based approaches are
orthogonal to our approach, as we are working on the source code directly. This
has the advantage that the developer does not need to learn a new (modeling)
language and an extra step of transformation (from model to source) is
eliminated.

\paragraph{SMT-based approaches.}

\textsc{Zeus}~\cite{kalra2018zeus} translates Solidity to LLVM bitcode and
employs existing verifiers such as \textsc{SeaHorn} and \textsc{Smack}. Besides
certain vulnerability patterns, it claims to have support for user-defined
properties to some extent. However, it is not publicly available for
comparison. \textsc{VeriSol}~\cite{lahiri2018formal} checks for conformance
between workflow policies and smart contract implementations on the Azure
blockchain. While the core of their method is a translation to Boogie (similar
to ours), it targets a specific problem limited in scope and does not yet
support features needed for typical smart contracts (see
Section~\ref{sec:results:annotated}). The Solidity compiler itself also includes
a built-in experimental SMT checker~\cite{leonardo2018smt}, which executes the
body of each function symbolically and checks for implicit specifications, such
as assertion failures, dead code and overflows. Their approach is however,
limited, by false overflow alarms and missing features (e.g., \codeintext{call},
\codeintext{revert}). Furthermore, it has no support for developer-supplied
specification beyond \codeintext{require} and \codeintext{assert} statements.
Some of the challenges they mention in their future work are solved by our
approach, including contract level invariants and the reduced number of false
overflow alarms.


\section{Conclusion}

We presented \tool, a tool for automated verification of
Solidity smart contracts based on modular program reasoning
and SMT solvers. Working at the source level, \tool allows
users to specify high-level properties such as contract
invariants, loop invariants, pre- and post-conditions and
assertions. \tool then discharges verification conditions
with SMT solvers to verify contract properties in a modular
and scalable way. The approach offers precise and scalable,
yet automated and user-friendly formal verification for
Solidity smart contracts. \tool can already be used on
real-world contracts and can effectively find bugs and prove
correctness of non-trivial properties with minimal user
effort.

%

\bibliographystyle{splncs04}
\bibliography{references}

\end{document}